\documentclass[11pt, reqno]{amsart}
\usepackage{graphicx,amsmath,amssymb,epsfig}
\usepackage{amsfonts}
\usepackage{harvard}
\usepackage{hyperref}

\hypersetup{
    colorlinks=true,
    urlcolor=blue,
    linkcolor=blue,
    citecolor=blue
}

\newtheorem{theorem}{Theorem}
\theoremstyle{plain}

\newtheorem{assumption}{Assumption}

\begin{document}
\title{Canonical Correlation and Assortative Matching: \\ A Remark}
\author{Arnaud Dupuy{\small $^{\S }$}}
\author{Alfred Galichon\textit{$^{\dag }$}}
\date{May 23, 2014. The authors thank two anonymous referees, the Editor
(Xavier D'Haultfoeuille), as well as Xavier Gabaix, Bernard\ Salani\'{e} and
Marko Tervi\"{o} for helpful comments. Galichon's research has received
funding from the European Research Council under the European Union's
Seventh Framework Programme (FP7/2007-2013) / ERC grant agreement no 313699,
and from FiME, Laboratoire de Finance des March\'{e}s de l'Energie. Dupuy
warmly thanks the Maastricht School of Management where part of this
research was performed.}

\begin{abstract}
In the context of the Beckerian theory of marriage, when men and women match on a single-dimensional index that is the weighted sum of their respective multivariate attributes, many papers in the literature have used linear canonical correlation, and related techniques, in order to estimate these weights. We argue that this estimation technique is inconsistent and suggest some solutions.
\end{abstract}

\maketitle

\noindent


{\footnotesize \textbf{JEL codes}: C78, D61, C13.\vskip50pt }

{\footnotesize {\small $^{\S }$}\textit{\ CEPS/INSTEAD, Maastricht School of
Management and IZA. Address: CEPS/INSTEAD, 3, avenue de la Fonte - L-4364
Esch-sur-Alzette, Luxembourg. Email: arnaud.dupuy@ceps.lu.} Tel:
+352585855551, Fax: +352585855700.
}

{\footnotesize \textit{$^{\dag }$}\textit{\ Sciences Po Paris, Department of
Economics, CEPR and IZA. Address: 27 rue Saint-Guillaume, 75007 Paris,
France. E-mail: alfred.galichon@sciencespo.fr.} Tel: +33(0)145498582, Fax:
+33(0)145497257.}

\newpage

\bigskip

\textbf{Introduction}. Who marries whom and why are questions that have
received tremendous attention by scientists from many different fields:
economics, sociology, psychology and biology. This literature shows that a
correlation between spouses' attributes exists for many attributes, i.e.
height, weight, education, earnings, wealth, religion, ethnicity,
personality traits to mention just a few. How many and which of these
attributes actually matter for the sorting of men and women? Up until
recently, by lack of a better methodology, the literature dealt with the
first question by simply assuming that men and women match on a
single-dimensional index that is the weighted sum of their respective
multivariate attributes. The second question was then dealt with using
linear Canonical Correlation, and related techniques, in order to estimate
the weights of the indices for men and women. This paper argues that this
estimation technique is inconsistent and suggest some solutions.

Since Becker's (1973) seminal contribution, the marriage market has been
predominantly modeled as a matching market with transferable utility. Men
and women are characterized by vectors of attributes denoted respectively $%
x\in \mathbb{R}^{d_{x}}$ for men and $y\in \mathbb{R}^{d_{y}}$ for women.
These vectors may incorporate various dimensions such as education, wealth,
health, physical attractiveness, personality traits, etc. It is assumed that
when a man with attributes $x$ and a woman with attributes $y$ form a pair,
they generate a surplus equal to $\Phi \left( x,y\right) $. This surplus is
shared endogenously between the two partners. Denoting $P$ and $Q$ the
respective probability distributions of attributes of married men and women,
it follows from the results of Shapley and Shubik (1972) that the stable
matching will maximize 
\begin{equation*}
\mathbb{E}\left[ \Phi \left( X,Y\right) \right]
\end{equation*}%
with respect to all joint distributions of $\left( X,Y\right) $ such that $%
X\sim P$ and $Y\sim Q$. For convenience, we assume that these distributions
are centered $\int xdP\left( x\right) =\int ydQ\left( y\right) =0$.

\bigskip

Becker went further in the analysis by assuming that sorting occurs on
single-dimensional \emph{ability indices} for men and women, say $\bar{x}$
and $\bar{y}$, which are constructed linearly with respect to the original
attributes%
\begin{equation*}
\bar{x}=\alpha ^{\prime }x\text{ and }\bar{y}=\beta ^{\prime }y
\end{equation*}%
where $\alpha \in \mathbb{R}^{d_{x}}$ and $\beta \in \mathbb{R}^{d_{y}}$ are
the weights according to which the various attributes enter the respective
indices. Following Becker (1973), assume that the matching surplus of
individuals of attributes $x$ and $y$, denoted $\Phi \left( x,y\right) $,
only depends on the indices $\bar{x}$ and $\bar{y}$ and takes the form 
\begin{equation*}
\Phi \left( x,y\right) =\phi \left( \alpha ^{\prime }x,\beta ^{\prime
}y\right)
\end{equation*}%
where $\phi $ is supermodular, that is $\partial _{\bar{x},\bar{y}}^{2}\phi
\left( \bar{x},\bar{y}\right) \geq 0$. As a result, the solution exhibits
positive assortative matching, that is, the equilibrium distribution of the
attributes across couples is represented by a joint random vector $\left(
X,Y\right) \sim \pi $ where $\alpha ^{\prime }X$ and $\beta ^{\prime }Y$ are 
\emph{comonotone}: the man at percentile $t$ in the distribution of $\alpha
^{\prime }X$ is matched with the woman at percentile $t$ in the distribution
of $\beta ^{\prime }Y$. In other words, denoting $F_{Z}$ the cumulative
distribution function of $Z$, we can state as the main assumption of this
note that:

\begin{assumption}
There are weights $\alpha $ and $\beta $ such that the indices $\alpha
^{\prime }X$ and $\beta ^{\prime }Y$ are comonotone, that is%
\begin{equation*}
F_{\beta ^{\prime }Y}\left( \beta ^{\prime }Y\right) =F_{\alpha ^{\prime
}X}\left( \alpha ^{\prime }X\right) .
\end{equation*}
\end{assumption}

If the cumulative distribution function $F_{\beta ^{\prime }Y}$ is
invertible, one may then write 
\begin{equation*}
\beta ^{\prime }Y=T\left( \alpha ^{\prime }X\right)
\end{equation*}%
where $T\left( z\right) =F_{\beta ^{\prime }Y}^{-1}\circ F_{\alpha ^{\prime
}X}\left( z\right) $ is a nondecreasing map; thus the ability index of a
woman is a nondecreasing function of that of the man she is matched with.

\bigskip

Given this specification and the observation of $\left( X,Y\right) \sim \pi $%
, one would like to estimate $\left( \alpha ,\beta \right) $. To this end,
Becker (1973) suggested (p. 834) to use Canonical Correlation Analysis, a
technique originally introduced by Hotelling (1936). This method consists in
determining the weights $\alpha ^{c}$ and $\beta ^{c}$ that maximize the
correlation between $\alpha ^{\prime }X$ and $\beta ^{\prime }Y$.\footnote{%
Since we are primarily interested about the consistency of Canonical
Correlation and related techniques, throughout this paper we assume that the
analyst has access to a sample of infinite size.} Formally, introducing the
following notations%
\begin{equation*}
\Sigma _{XY}=\mathbb{E}_{\pi }\left[ XY^{\prime }\right] ,~\Sigma _{X}=%
\mathbb{E}_{\pi }\left[ XX^{\prime }\right] ,~\Sigma _{Y}=\mathbb{E}_{\pi }%
\left[ YY^{\prime }\right] ,
\end{equation*}%
Canonical Correlation consists in defining $\alpha ^{c}$ and $\beta ^{c}$ as
the maximizers of the correlation of $\alpha ^{\prime }X$ and $\beta
^{\prime }Y$ over all possible vectors of weights $\alpha $ and $\beta $.
The problem therefore consists in solving the following program%
\begin{eqnarray}
&&\max_{\alpha \in \mathbb{R}^{d_{x}},\beta \in \mathbb{R}^{d_{y}}}\alpha
^{\prime }\Sigma _{XY}\beta  \label{cancorr} \\
&&s.t.\text{ }\alpha ^{\prime }\Sigma _{X}\alpha =1\text{ and }\beta
^{\prime }\Sigma _{Y}\beta =1  \notag
\end{eqnarray}%
whose value at optimum is in general less or equal than one.

\bigskip

In the applied literature, $\alpha $ and $\beta $ are frequently estimated
by multivariate Ordinary Least Squares (OLS) regression. It is worth
remarking that this is closely related, but not quite identical to,
Canonical Correlation. Consider the following OLS regression%
\begin{equation*}
Y_{1}=\alpha ^{\prime }X-\beta _{-1}^{^{\prime }}Y_{-1}+\varepsilon
\end{equation*}%
where $\varepsilon $ is an error term, $Y_{1}$ is the top element of $Y$,
and $Y_{-1}$ the vector of the remaining entries. Let $\alpha ^{o}$ and $%
\beta _{-1}^{o}$ be the coefficients obtained from OLS. Introducing $\beta
^{o}=\left( 1~\beta _{-1}^{o\prime }\right) ^{\prime }$, it is easy to show
that $\left( \alpha ^{o},\beta ^{o}\right) $ solves the program%
\begin{eqnarray*}
&&\max_{\alpha \in \mathbb{R}^{d_{x}},\beta \in \mathbb{R}^{d_{y}}}\alpha
^{\prime }\Sigma _{XY}\beta \\
&&s.t.\text{ }\alpha ^{\prime }\Sigma _{X}\alpha =A\text{ and }\beta
^{\prime }\Sigma _{Y}\beta =B\text{ and }\beta _{1}=1.
\end{eqnarray*}%
where $A=\alpha ^{o\prime }\Sigma _{X}\alpha ^{o}$ and $B=\beta
^{o}{}^{\prime }\Sigma _{Y}\beta ^{o}$. Without the constraint $\beta _{1}=1$%
, this would yield the same solutions (up to some rescaling of $\alpha $ and 
$\beta $) as the solutions given by Canonical Correlation. In general, the
solutions differ due to this constraint. Even though the OLS technique is
better known and more immediately accessible to practitioners, it
artificially breaks down symmetry between variables by singling out the role
of $Y_{1}$. Note that in the case where $Y$ is univariate ($d_{y}=1$) the
constraint $\beta _{1}=1$ has no bite, and the two solutions coincide
(again, up to rescaling).

\bigskip

Many papers have used Canonical Correlation or OLS techniques to estimate $%
\alpha $ and $\beta $. Notable examples of the application of Canonical
Correlation on the marriage market are Suen and Lui (1999), Gautier et al.
(2005) and Taubman (2006). Many papers have applied OLS techniques to study
assortative mating when faced with multiple dimensions, see Kalmijn (1998)
for a survey of this literature. A notable example of such applications of
OLS is the extensive literature on the effect of a wife's education on her
husband's earnings: see among others Benham (1974), Scully (1979), Wong
(1986), Lam and Schoeni (1993, 1994), and Jepsen (2005).

\bigskip

\textbf{The consistency problem}. A crucial question is whether the
Canonical Correlation method is consistent, namely whether $\left( \alpha
^{c},\beta ^{c}\right) =\left( \alpha ,\beta \right) $. It turns out that
the answer is yes in the case of Gaussian marginal distributions $P$ and $Q$%
, but no in more general cases as we shall now explain. We now state our
result. The main statement, part (ii) of the theorem, is proven using a
counterexample.

\begin{theorem}[(In-)Consistency of Canonical Correlation]
\label{thm:canonicalCorrel}The following holds:

(i) If $P\ $and $Q$ are Gaussian distributions, then the Canonical
Correlation is consistent in the sense that%
\begin{equation*}
\left( \alpha ^{c},\beta ^{c}\right) =\left( \alpha ,\beta \right) .
\end{equation*}%
(ii) In general, Canonical Correlation is not consistent.
\end{theorem}

\bigskip

\begin{proof}
(i) When $P=N\left( 0,\Sigma _{X}\right) $ and $Q=N\left( 0,\Sigma
_{Y}\right) $, with $\alpha ,\beta \neq 0$ two vectors of weights, then%
\begin{equation*}
\max_{X\sim P,Y\sim Q}\mathbb{E}\left[ \alpha ^{\prime }XY^{\prime }\beta %
\right] =\sqrt{\alpha ^{\prime }\Sigma _{X}\alpha }\sqrt{\beta ^{\prime
}\Sigma _{Y}\beta },
\end{equation*}%
where the optimization is over the set of random vectors $\left( X,Y\right) $
with fixed marginal distributions $P$ and $Q$. Thus, for $\left(
X\,,Y\right) $ solution of the above problem, the correlation between $%
\alpha ^{\prime }X$ and $\beta ^{\prime }Y$ is one. Indeed, the optimal $%
\left( X,Y\right) $ is such that%
\begin{equation*}
\beta ^{\prime }Y=\sqrt{\frac{\beta ^{\prime }\Sigma _{Y}\beta }{\alpha
^{\prime }\Sigma _{X}\alpha }}\alpha ^{\prime }X.
\end{equation*}

The result is immediate: for the optimal $\left( X,Y\right) $, the
correlation between $\alpha ^{\prime }X$ and $\beta ^{\prime }Y$ is one and
since this is the maximal value of Program (\ref{cancorr}), it follows that $%
\left( \alpha ,\beta \right) =\left( \alpha ^{c},\beta ^{c}\right) $.

\bigskip

(ii) However, when $P$ and $Q$ fail to be Gaussian, the Canonical
Correlation estimator $\left( \alpha ^{c},\beta ^{c}\right) $ differs from
the true parameters $\left( \alpha ,\beta \right) $ in general. Consider the
following example. Let $P$ be the distribution of $\left( X_{1},X_{2}\right) 
$ where $X_{1}$ is independent of $X_{2}$ and $V\left( X_{1}\right) =V\left(
X_{2}\right) =1$. Let $Q$ be the distribution of $Y$. Provided that the
surplus function satisfies $\Phi \left( x,y\right) =\phi \left( \alpha
^{\prime }x,\beta ^{\prime }y\right) $ such that sorting is unidimensional,
optimal matching yields: $Y=\frac{T\left( \alpha _{1}X_{1}+\alpha
_{2}X_{2}\right) }{\beta }$ where $T:=F_{\alpha ^{^{\prime }}X}^{-1}\left(
F_{\beta Y}\left( .\right) \right) $ and $F_{\gamma ^{^{\prime }}Z}$ denotes
the c.d.f. of $\gamma ^{^{\prime }}Z$. Note that the mapping function $T$
depends on $P$, $Q$, $\alpha $ and $\beta $. In this setting, the Canonical
Correlation estimator $\left( \alpha _{1}^{c},\alpha _{2}^{c}\right) $ of $%
\left( \alpha _{1},\alpha _{2}\right) $ solves%
\begin{eqnarray*}
&&\max_{\alpha _{1},\alpha _{2}}\alpha _{1}cov\left( X_{1},Y\right) +\alpha
_{2}cov\left( X_{2},Y\right)  \\
&&s.t.~\alpha _{1}^{2}+\alpha _{2}^{2}=1
\end{eqnarray*}

whose solution is%
\begin{equation}
\frac{\alpha _{1}^{c}}{\alpha _{2}^{c}}=\frac{cov\left( X_{1}Y\right) }{%
cov(X_{2}Y)}.  \label{eqSol}
\end{equation}%
In such an economy, data on \textquotedblleft couples\textquotedblright\ are
such that $Y=\frac{T\left( \alpha _{1}X_{1}+\alpha _{2}X_{2}\right) }{\beta }
$ for all $X$. Replacing $Y$ by its expression in terms of $X$ in the right
hand side of eq. \ref{eqSol} yields%
\begin{equation*}
\frac{\alpha _{1}^{c}}{\alpha _{2}^{c}}=\frac{cov\left( X_{1}T\left( \alpha
_{1}X_{1}+\alpha _{2}X_{2}\right) \right) }{cov(X_{2}T\left( \alpha
_{1}X_{1}+\alpha _{2}X_{2}\right) )}.
\end{equation*}%
It follows that the Canonical Correlation estimator is consistent if and
only if%
\begin{equation*}
\frac{\alpha _{1}^{c}}{\alpha _{2}^{c}}=\frac{\alpha _{1}}{\alpha _{2}}
\end{equation*}%
that is if and only if%
\begin{equation}
\frac{\alpha _{1}}{\alpha _{2}}=\frac{cov\left( X_{1}T\left( \alpha
_{1}X_{1}+\alpha _{2}X_{2}\right) \right) }{cov(X_{2}T\left( \alpha
_{1}X_{1}+\alpha _{2}X_{2}\right) )}.  \label{eqCond}
\end{equation}%
It is easy to see that this condition will be satisfied when ever $T$ is
linear (with constant $a$ and slope $b$), a case that arises for instance
when $P$ and $Q$ are Gaussian as in i), for then one has 
\begin{equation*}
\frac{cov\left( X_{1}\left( a+b\alpha _{1}X_{1}+b\alpha _{2}X_{2}\right)
\right) }{cov(X_{2}\left( a+b\alpha _{1}X_{1}+b\alpha _{2}X_{2}\right) )}=%
\frac{\alpha _{1}}{\alpha _{2}}.
\end{equation*}%
However, as soon as $T$ is nonlinear, there are no reasons to expect that $T$
will satisfy condition \ref{eqCond}. For instance, let $P$ be the
distribution of $\left( X_{1},X_{2}\right) $ where $X_{1}$ takes value $1$
with probability $1/2$ and $-1$ with probability $1/2$, and $X_{2}$ is
exponentially distributed with parameter 1 and independent of $X_{1}$. Let $G
$ be the c.d.f. of $X_{2}$, so that $G\left( z\right) =1-\exp \left(
-z\right) $. Let $Q=\mathcal{U}\left( \left[ 0,1\right] \right) $. Set $%
\alpha _{1}=\alpha _{2}=1/\sqrt{2}$, so that $\hat{X}=\frac{X_{1}+X_{2}}{%
\sqrt{2}}$. Hence the optimal coupling $\left( \hat{X},\hat{Y}\right) $ is
such that $\hat{Y}=F_{\hat{X}}\left( \hat{X}\right) $ where $F_{\hat{X}%
}\left( .\right) $ is the c.d.f. of $\hat{X}$, which is expressed as%
\begin{equation*}
F_{\hat{X}}\left( x\right) =\frac{1}{2}\left( G\left( x\sqrt{2}+1\right)
+G\left( x\sqrt{2}-1\right) \right) .
\end{equation*}

Clearly, in this example $T:=F_{\hat{X}}$ is not linear in this case such
that one should not expect Canonical Correlation to be consistent. And,
indeed, one has%
\begin{equation*}
\hat{Y}=\left\{ 
\begin{array}{c}
\frac{1}{2}\left( G\left( X_{2}\right) +G\left( X_{2}-2\right) \right) \text{
if }X_{1}=-1 \\ 
\frac{1}{2}\left( G\left( X_{2}+2\right) +G\left( X_{2}\right) \right) \text{
if }X_{1}=1,%
\end{array}%
\right.
\end{equation*}%
and a calculation shows that%
\begin{equation*}
cov\left( X_{1},\hat{Y}\right) =\frac{\mathbb{E}G\left( X_{2}+2\right) -%
\mathbb{E}G\left( X_{2}-2\right) }{4}
\end{equation*}%
and as $\mathbb{E}G\left( X_{2}+2\right) =1-e^{-2}/2$ and $\mathbb{E}G\left(
X_{2}-2\right) =e^{-2}/2$, we get%
\begin{equation}
cov\left( X_{1},\hat{Y}\right) =\frac{1}{4}\left( 1-e^{-2}\right) .
\label{cov1}
\end{equation}%
Similarly,%
\begin{equation*}
\mathbb{E}\left[ X_{2}\hat{Y}\right] =\frac{1}{4}\mathbb{E}\left[
X_{2}G\left( X_{2}-2\right) \right] +\frac{1}{4}\mathbb{E}\left[
X_{2}G\left( X_{2}+2\right) \right] +\frac{1}{2}\mathbb{E}\left[
X_{2}G\left( X_{2}\right) \right]
\end{equation*}%
and using the fact that $\mathbb{E}\left[ X_{2}G\left( X_{2}-2\right) \right]
=7e^{-2}/4$, that $\mathbb{E}\left[ X_{2}G\left( X_{2}+2\right) \right]
=1-e^{-2}/4$, and that $\mathbb{E}\left[ X_{2}G\left( X_{2}\right) \right]
=3/4$, we get $\mathbb{E}\left[ X_{2}\hat{Y}\right] =\left( 3e^{-2}+5\right)
/8$, hence, as $\mathbb{E}\left[ X_{2}\right] \mathbb{E}\left[ \hat{Y}\right]
=1/2$, one obtains 
\begin{equation}
cov\left( X_{2},\hat{Y}\right) =\frac{3e^{-2}+1}{8}.  \label{cov2}
\end{equation}

Using (\ref{cov1}) and (\ref{cov2}), this becomes%
\begin{equation*}
\frac{\alpha _{2}^{c}}{\alpha _{1}^{c}}=\frac{3+e^{2}}{2e^{2}-2}\neq \frac{%
\alpha _{2}}{\alpha _{1}}=1.
\end{equation*}%
Therefore the Canonical Correlation estimator is not consistent in this
example.
\end{proof}

Note that the example in part (ii) of the proof also shows that OLS is
inconsistent. In this example the dimension of $Y$ is one, so that OLS and
Canonical Correlation yield the same estimators of $\alpha $ and $\beta $.
The above example has nothing pathological and implies that estimators of $%
\left( \alpha ,\beta \right) $ based on Canonical Correlation face the risk
of being biased as soon as the marginal distributions are not Gaussian or
such that the mapping function $T:=F_{\alpha ^{^{\prime }}X}^{-1}\left(
F_{\beta ^{\prime }Y}\left( .\right) \right) $ is not linear.

\bigskip

\textbf{Final remarks}. The problem discussed in this paper obviously raises
the question: how can we replace Canonical Correlation by a technique that
is consistent? One first proposal, as suggested in Tervi\"{o} (2003, p. 83),
is to look for $\alpha $ and $\beta $ that maximize Spearman's rank
correlation between $\alpha ^{\prime }X$ and $\beta ^{\prime }Y$. In other
words, look for 
\begin{eqnarray*}
&&\max_{\alpha \in \mathbb{R}^{d_{x}},\beta \in \mathbb{R}^{d_{y}}}\mathbb{E}%
\left[ F_{\alpha ^{\prime }X}\left( \alpha ^{\prime }X\right) F_{\beta
^{\prime }Y}\left( \beta ^{\prime }Y\right) \right] \\
&&s.t.\text{ }\alpha ^{\prime }\Sigma _{X}\alpha =1\text{ and }\beta
^{\prime }\Sigma \beta =1.
\end{eqnarray*}%
where we recall that $F_{\alpha ^{\prime }X}$ stands for the c.d.f. of $%
\alpha ^{\prime }X$. The value of this program cannot exceed 1/3 and, when
the distributions of $X$ and $Y$ are continuous, it is equal to 1/3 when $%
\alpha ^{\prime }X$ and $\beta ^{\prime }Y$ are comonotone. However the
objective function, which can be rewritten as%
\begin{equation*}
\int \Pr \left( \max \left( \alpha ^{\prime }\left( x-X\right) ,\beta
^{\prime }\left( y-Y\right) \right) \leq 0\right) dF_{X}\left( x\right)
dF_{Y}\left( y\right) ,
\end{equation*}%
has no reason to be convex with respect to $\alpha $ and $\beta $, so global
optimization techniques may be needed. Also, this technique, just as
Canonical Correlation, does not deal with any kind of unobserved
heterogeneity. To remedy this drawback, two solutions have very recently
been proposed.

The first solution is justified if one is willing to assume that sorting
occurs on a single index of attractiveness. This strategy, developed by
Chiappori et al. (2012), consists in estimating the conditional expectations 
$\mathbb{E}\left[ Y_{k}|X=x\right] $, which, if the sorting actually occurs
on a single-index, should be a deterministic function of $\alpha ^{\prime }X$%
. Hence the weight vector $\alpha $ is identified up to a constant by the
marginal rates of substitutions%
\begin{equation*}
\frac{\alpha _{i}}{\alpha _{j}}=\frac{\partial \mathbb{E}\left[ Y_{k}|X=x%
\right] /\partial x_{i}}{\partial \mathbb{E}\left[ Y_{k}|X=x\right]
/\partial x_{j}}.
\end{equation*}

Moving outside of single-dimensional indices, Dupuy and Galichon (2014) have
introduced a technique they call \textquotedblleft saliency
analysis\textquotedblright , which allows to infer the \emph{number of
dimensions} on which sorting occurs, and estimate the corresponding
(possibly multiple) \emph{indices of attractiveness} that determine this
sorting.

The idea is to estimate $A$ in the quadratic specification for the surplus
function 
\begin{equation*}
\Phi \left( x,y\right) =x^{\prime }Ay,
\end{equation*}%
applying for instance the procedure depicted in Dupuy and Galichon (2014),
and using a singular value decomposition to test whether the dimension of $A$
is e.g. one, in which case $A=\alpha \beta ^{\prime }$. This provides a
consistent estimation of $\alpha $ and $\beta $. Note however, that the
units of the parameters of the affinity matrix reflect the units in which $X$
and $Y$ are measured. For our method to be robust to changes in measurement
units, we need to normalize the attributes in $X$ and $Y$. By performing the
Singular Value Decomposition on the affinity matrix associated with the
normalized attributes, we ensure that the loadings of the indices of mutual
attractiveness are independent of the choice of measurement units. For the
sake of notation and compactness we herewith simply assume that $X$ and $Y$
have been rescaled such that all attributes are of variance $1$.

Performing a singular value decomposition of $A$ yields%
\begin{equation*}
A=U^{\prime }\Lambda V,
\end{equation*}%
where the diagonal matrix $\Lambda $ has nonincreasing elements $\left(
\lambda _{1},...,\lambda _{d}\right) $ called singular values, $d=\min
\left( d_{x},d_{y}\right) $ on its diagonal. By construction, $U$ and $V$
are orthogonal matrices.

One can then define vectors of \emph{indices of mutual attractiveness} by
constructing%
\begin{equation*}
\tilde{X}=UX\text{ and }\tilde{Y}=VY,
\end{equation*}%
where each index is a weighted sum of the attributes in $X$ and $Y$
respectively.

Denote $A^{\tilde{X}\tilde{Y}}$ the affinity matrix corresponding to the
vectors of characteristics $\tilde{X}$ and $\tilde{Y}$. Dupuy and Galichon
(2014) have shown that in fact $A^{\tilde{X}\tilde{Y}}=\Lambda $, and as a
result 
\begin{equation*}
\Phi \left( x,y\right)
=\sum_{i=1}^{d_{x}}\sum_{j=1}^{d_{y}}A_{ij}x_{i}y_{j}=\sum_{i=1}^{d}\lambda
_{i}\tilde{x}_{i}\tilde{y}_{i}.
\end{equation*}

The weights of each index of mutual attractiveness constructed by Saliency
Analysis can be read on the associated row of $U$ for men and $V$ for women
whereas the share of the matching utility of couples explained by the $%
i^{th} $ pair of indices is given by $\lambda _{i}/(\sum_{i}\lambda _{i})$.
Saliency Analysis answers two important questions: how many and which
attributes matter for the sorting of men and women on the marriage market.
Intuitively, the number of non zero singular values indicates the number of
indices that matter for the sorting problem and the parameters of $U$ and $V$
indicate which attributes matter in each index of men and each index of
women. If there is only one non zero singular value, then sorting occurs on
a single index whose weights are given by the first row of $U$ for men and $%
V $ for women and correspond to $\alpha $ and $\beta $ respectively.

\end{document}